\documentclass[prd,aps,showpacs,nofootinbib,superscriptaddress,notitlepage,10pt,tightenlines]{revtex4-1}
\usepackage{epsfig}
\usepackage{amssymb}
\usepackage{amsmath}
\usepackage{graphicx}
\usepackage{psfrag}

\begin{document}

\title{On the small-$x$ evolution of the color quadrupole and the Weizs\"{a}cker-Williams gluon distribution}
\author{Fabio Dominguez}
\author{A. H. Mueller}
\affiliation{Department of Physics, Columbia University, New York,
NY, 10027, USA}
\author{St\'ephane Munier}
\affiliation{Centre de physique th{\'e}orique, \'Ecole Polytechnique,
CNRS, Palaiseau, France}
\author{Bo-Wen Xiao}
\affiliation{Department of Physics, Pennsylvania State University, University Park, PA 16802, USA}

\begin{abstract}
Color quadrupoles have been found to be important in the proper description of observables sensitive to the small-$x$ regime in nuclei as well as in the operator definition of the Weizs\"{a}cker-Williams gluon distribution. In this paper, we derive the small-$x$ evolution equation of the quadrupole and the Weizs\"{a}cker-Williams gluon distribution without taking large $N_c$ limit and study the properties of the equation in both dilute and saturation regime. We find that the quadrupole evolution follows the BFKL evolution in the dilute regime and then saturates in the dense region due to nonlinear terms. This leads us to conclude that the Weizs\"{a}cker-Williams gluon distribution should obey the same geometrical behavior as the dipole gluon distribution as found in the inclusive DIS measurement. 
\end{abstract}
\pacs{24.85.+p, 12.38.Bx, 12.39.St, 13.88.+e}
\maketitle

\section{Introduction}

Recent studies of two-particle production processes \cite{Marquet:2007vb, Albacete:2010pg, JalilianMarian:2004da, Dominguez:2010xd, Dominguez:2011wm} where a dilute probe scatters from a dense target have shown that in order to understand saturation phenomena and the corresponding small-$x$ dynamics it is necessary to go beyond color dipoles and compute the appropriate evolution equations for color quadrupoles as well. In particular, the measurement of two-particle correlations in d+Au collisions \cite{Braidot:2010ig,Adare:2011sc} has attracted a fair amount of attention due to the observed suppression of the far-side peak as the rapidity of the observed particles is increased. So far, the only satisfactory explanation for this phenomenon, showing suppression for forward rapidities and not in the central rapidity region, is the calculation presented in \cite{Albacete:2010pg} which is performed in a Color Glass Condensate (CGC) framework. Even though the results of \cite{Albacete:2010pg} are shown to agree with the current data, it has been argued \cite{Dominguez:2011wm,Dumitru:2010ak} that the approximation used to implement the small-$x$ evolution of the relevant correlators in terms of only color dipoles is not correct even in the large-$N_c$ limit. Based on the study of two-particle production processes in DIS and pA collisions \cite{Dominguez:2011wm,JalilianMarian:2004da} it is now believed that for scattering processes involving a dilute probe interacting with a dense target can be described in terms of only color dipoles and color quadrupoles in the large-$N_c$ limit.

Color quadrupoles have also been shown to be relevant when unintegrated gluon distributions are considered in the small-$x$ regime. In small-$x$ physics, two different unintegrated gluon distributions \cite{Catani:1990eg, Collins:1991ty} (also called transverse momentum dependent gluon distributions), namely the Weizs\"{a}cker-Williams gluon distribution $xG^{(1)}$ \cite{Kovchegov:1998bi,McLerran:1998nk} and the dipole gluon distribution $xG^{(2)}$, have been found and widely used in the literature. The Weizs\"{a}cker-Williams gluon distribution, known as the conventional gluon distribution, gives the Fock space number density of gluons inside dense hadrons in light-cone gauge. The dipole gluon distribution, defined via the Fourier transform of the color dipole amplitude, has been studied thoroughly since it appears in many physical processes \cite{Iancu:2003xm, Kharzeev:2003wz}. This dipole gluon distribution can be probed directly in photon-jet correlations measurement in pA collisions. Recently, studies \cite{Dominguez:2010xd, Dominguez:2011wm} on the Weizs\"{a}cker-Williams gluon distribution indicate that it can be directly measured in DIS dijet production and its operator definition is related to quadrupoles instead of normal color dipoles. Other more complicated dijet processes in pA collisions (e.g., $qg$ or $gg$ dijets) involve both of these gluon distributions through convolution in transverse momentum space. 


In terms of the operator definitions of the Weizs\"{a}cker-Williams gluon distribution $xG^{(1)}$ and the dipole gluon distribution $xG^{(2)}$,the evolution of $xG^{(2)}$ appears to be given by the evolution of the dipole scattering amplitude which obeys the Balitsky-Kovchegov equation \cite{Balitsky:1995ub+X,Kovchegov:1999yj}, while $xG^{(1)}$ is governed by the evolution of both dipole and quadrupole scattering amplitudes. For more complicated dijet processes in pA collisions \cite{Dumitru:2010ak}, one needs to use both the dipole and the quadrupole evolution equations since the cross sections of those processes involve both gluon distributions and their combinations. Knowing both dipole and quadrupole forward scattering amplitudes on a target allows one to understand their small-$x$ behavior and thus compute physical observables in high energy scatterings in large $N_c$ limit. This is necessary to describe dijet processes in both DIS and pA collisions systematically. 
 
The objective of this paper is to derive the small-$x$ evolution equation of the quadrupole and the Weizs\"{a}cker-Williams gluon distribution without taking the large $N_c$ limit, and study their small-$x$ behavior. As far as the quadrupole amplitude $Q$ is concerned, it seems that qualitatively it is understood quite well. In the leading twist limit when the density is low, it obeys a BFKL type equation. When all the sizes of the quadrupole are comparable to the scale at which unitarity sets in, namely $\Delta x_i \sim \frac{1}{Q_s(Y)}$($Q_s(Y)$ is the so-called saturation
momentum, which depends on the rapidity $Y$, and whose inverse gives
the order of magnitude of the size above which color-neutral objects 
are absorbed.), the quadrupole amplitude evolves towards the stable fixed point $Q=0$ and the corresponding Weizs\"{a}cker-Williams gluon distribution $xG^{(1)}$ starts to saturate. 

The rest of the paper is organized as follows. In Sec. II, we carry out the derivations for the evolution equation of the quadrupole amplitudes and the Weizs\"{a}cker-Williams gluon distribution.  We investigate the weak interaction limit of these equations in Sec. III. Sec. IV is devoted to the discussion on the unitarity limit and fixed points of the quadrupole evolution equation. The summary and further discussions are given in Sec. V. 

\section{The evolution equation of the quadrupole}
In general, the JIMWLK evolution \cite{Jalilian-Marian:1997jx+X, Ferreiro:2001qy} of an arbitrary operator is given by 
\begin{equation}
\frac{\partial\langle\mathcal{O}\rangle_Y}{\partial Y}=\frac{1}{2}\int d^2 u_{\perp} \int d^2 v_{\perp} 
\left\langle \frac{\delta}{\delta \alpha^a_{u_{\perp}}}\eta_{uv}^{ab}\frac{\delta}{\delta \alpha^b_{v_{\perp}}}\mathcal{O}\right\rangle_Y, \label{JH}
\end{equation}
with 
\begin{equation}
\eta_{uv}^{ab}=\frac{1}{\pi}\int \frac{d^2 z_{\perp}}{(2\pi)^2}\mathcal{K}\left(u_{\perp},v_{\perp},z_{\perp}\right)\left[1+\tilde{V}^{\dagger}_{u}\tilde{V}_{v}-\tilde{V}^{\dagger}_{u}\tilde{V}_{z}-\tilde{V}^{\dagger}_{z}\tilde{V}_{v}\right]^{ab}, \label{JK}
\end{equation}
where $\mathcal{K}\left(u_{\perp},v_{\perp},z_{\perp}\right)=\frac{(u_{\perp}-z_{\perp})\cdot(v_{\perp}-z_{\perp})}{(u_{\perp}-z_{\perp})^2(v_{\perp}-z_{\perp})^2}$ and the $\tilde{V}$ represents the Wilson line in the adjoint representation.  

If one sets $\mathcal{O}=\frac{1}{N_c}\left\langle\textrm{Tr}\left(U(x_{\perp})U^{\dagger}(y_{\perp})\right)\right\rangle_{Y}$ with $U(x_{\perp})$ being the Wilson line in the fundamental representation, one can easily reproduce the well-known evolution equation for the dipole amplitude,
\begin{eqnarray}
\frac{\partial}{\partial Y}\left\langle\textrm{Tr}\left[U(x)U^{\dagger}(y)\right]\right\rangle_{Y}&=&-\frac{\alpha_s N_c}{2\pi^2}\int d^2 z_{\perp} \frac{(x_{\perp}-y_{\perp})^2}{(x_{\perp}-z_{\perp})^2(z_{\perp}-y_{\perp})^2} \nonumber \\
&&\times \left\{ \left\langle\textrm{Tr}\left[U(x)U^{\dagger}(y)\right]\right\rangle_{Y}-\frac{1}{N_c}\left\langle\textrm{Tr}\left[U(x)U^{\dagger}(z)\right]\textrm{Tr}\left[U(z)U^{\dagger}(y)\right]\right\rangle_{Y}\right\}.\label{bk}
\end{eqnarray}
This is not a closed equation for the dipole amplitude since the last term involves a correlator of four Wilson lines. Using a mean field approximation for a large nucleus and in the large-$N_c$ limit, this correlator can be factorized as a product of two dipole amplitudes yielding the well-known Balitsky-Kovchegov equation. The Balitsky-Kovchegov equation has a probabilistic interpretation in terms of dipole splittings with the probability of one dipole ($x_\perp,y_\perp$) splitting into two new dipoles ($x_\perp,z_\perp)$ and $(z_\perp,y_\perp$) being $P_{d\to dd} =\frac{\alpha_sN_c}{2\pi^2}\frac{d^2 z_{\perp}(x_{\perp}-y_{\perp})^2}{(x_{\perp}-z_{\perp})^2(z_{\perp}-y_{\perp})^2} $.

Here we would like to derive the evolution equation of quadrupoles by using the JIMWLK Hamiltonian method without taking the large $N_c$ limit. By setting $\mathcal{O}=\frac{1}{N_c}\left\langle\textrm{Tr}\left(U(x_{1})U^{\dagger}(x^{\prime}_{1})U(x_{2})U^{\dagger}(x^{\prime}_{2})\right)\right\rangle_{Y}$ and using the above Hamiltonian, we get (see the derivation in Appendix.~\ref{derivation}) 
\begin{eqnarray}
&&\frac{\partial}{\partial Y}\left\langle\textrm{Tr}\left[U(x_{1})U^{\dagger}(x^{\prime}_{1})U(x_{2})U^{\dagger}(x^{\prime}_{2})\right]\right\rangle_{Y} \nonumber \\
&=&-\frac{\alpha_s N_c}{(2\pi)^2}\int d^2 z_{\perp} \mathcal{K}_1(x_1, x_1^{\prime},x_2, x_2^{\prime};z)\left\langle\textrm{Tr}\left[U(x_{1})U^{\dagger}(x^{\prime}_{1})U(x_{2})U^{\dagger}(x^{\prime}_{2})\right]\right\rangle_{Y}\nonumber \\
&&+\frac{\alpha_s N_c}{(2\pi)^2}\int d^2 z_{\perp} \mathcal{A}(x_1, x_1^{\prime},x_2, x_2^{\prime};z)
\frac{1}{N_c}\left\langle\textrm{Tr}\left[U^{\dagger}(x^{\prime}_{1})U(x_{2})\right]\textrm{Tr}\left[U^{\dagger}(x^{\prime}_{2})U(x_{1})\right]\right\rangle_{Y} \nonumber \\
&&+\frac{\alpha_s N_c}{(2\pi)^2}\int d^2 z_{\perp} \mathcal{B}(x_1, x_1^{\prime},x_2, x_2^{\prime};z)
\frac{1}{N_c}\left\langle\textrm{Tr}\left[U(x_{1})U^{\dagger}(x^{\prime}_{1})\right]\textrm{Tr}\left[U(x_{2})U^{\dagger}(x^{\prime}_{2})\right]\right\rangle_{Y} \nonumber \\
&&+\frac{\alpha_s N_c}{(2\pi)^2}\int d^2 z_{\perp} \mathcal{K}_2(x_1; x_1^{\prime}, x_2^{\prime};z)\frac{1}{N_c}\left\langle\textrm{Tr}\left[U(x_{1})U^{\dagger}(z)\right]\textrm{Tr}\left[U(z)U^{\dagger}(x^{\prime}_{1})U(x_{2})U^{\dagger}(x^{\prime}_{2})\right]\right\rangle_{Y}\nonumber \\
&&+\frac{\alpha_s N_c}{(2\pi)^2}\int d^2 z_{\perp} \mathcal{K}_2(x_1^{\prime}; x_1, x_2;z)\frac{1}{N_c}\left\langle\textrm{Tr}\left[U(z)U^{\dagger}(x^{\prime}_1)\right]\textrm{Tr}\left[U(x_1)U^{\dagger}(z)U(x_{2})U^{\dagger}(x^{\prime}_{2})\right]\right\rangle_{Y}\nonumber \\
&&+\frac{\alpha_s N_c}{(2\pi)^2}\int d^2 z_{\perp} \mathcal{K}_2(x_2; x_1^{\prime}, x_2^{\prime};z)\frac{1}{N_c}\left\langle\textrm{Tr}\left[U(x_2)U^{\dagger}(z)\right]\textrm{Tr}\left[U(x_1)U^{\dagger}(x_1^{\prime})U(z)U^{\dagger}(x^{\prime}_{2})\right]\right\rangle_{Y}\nonumber \\
&&+\frac{\alpha_s N_c}{(2\pi)^2}\int d^2 z_{\perp} \mathcal{K}_2(x_2^{\prime}; x_1, x_2;z)\frac{1}{N_c}\left\langle\textrm{Tr}\left[U(z)U^{\dagger}(x^{\prime}_2)\right]\textrm{Tr}\left[U(x_1)U^{\dagger}(x_1^{\prime})U(x_{2})U^{\dagger}(z)\right]\right\rangle_{Y},\label{quad}
\end{eqnarray}
where 
\begin{eqnarray}
\mathcal{K}_1(x_1, x_1^{\prime},x_2, x_2^{\prime};z)&=&\frac{(x^{\prime}_1-x_1)^2}{(x^{\prime}_1-z)^2(z-x_1)^2}+\frac{(x^{\prime}_2-x_2)^2}{(x^{\prime}_2-z)^2(z-x_2)^2}+\frac{(x^{\prime}_1-x_2)^2}{(x^{\prime}_1-z)^2(z-x_2)^2}+\frac{(x^{\prime}_2-x_1)^2}{(x^{\prime}_2-z)^2(z-x_1)^2}, \\
\mathcal{A}(x_1, x_1^{\prime},x_2, x_2^{\prime};z)&=&\frac{(x^{\prime}_1-x_2^{\prime})^2}{(x^{\prime}_1-z)^2(z-x_2^{\prime})^2}+\frac{(x_1-x_2)^2}{(x_1-z)^2(z-x_2)^2}-\frac{(x^{\prime}_1-x_2)^2}{(x^{\prime}_1-z)^2(z-x_2)^2}-\frac{(x^{\prime}_2-x_1)^2}{(x^{\prime}_2-z)^2(z-x_1)^2},\\
\mathcal{B}(x_1, x_1^{\prime},x_2, x_2^{\prime};z)&=&\frac{(x^{\prime}_1-x_2^{\prime})^2}{(x^{\prime}_1-z)^2(z-x_2^{\prime})^2}+\frac{(x_1-x_2)^2}{(x_1-z)^2(z-x_2)^2}-\frac{(x^{\prime}_2-x_2)^2}{(x^{\prime}_2-z)^2(z-x_2)^2}-\frac{(x^{\prime}_1-x_1)^2}{(x^{\prime}_1-z)^2(z-x_1)^2},\\
\mathcal{K}_2(x_1; x_1^{\prime},x_2^{\prime};z)&=&\frac{(x_1-x_1^{\prime})^2}{(x_1-z)^2(z-x_1^{\prime})^2}+\frac{(x_1-x_2^{\prime})^2}{(x_1-z)^2(z-x_2^{\prime})^2}-\frac{(x^{\prime}_1-x_2^{\prime})^2}{(x^{\prime}_1-z)^2(z-x_2^{\prime})^2}.\label{k2d} 
\end{eqnarray}
We note that the kernels satisfy the sum rule
\begin{equation}
 \mathcal{K}_1=\mathcal{A}+\mathcal{B}+\sum^4 \mathcal{K}_2 \label{sum},
\end{equation}
where the sum in Eq.~(\ref{sum}) goes over the missing coordinate in $\mathcal{K}_2$ which in Eq.~(\ref{k2d}) is $x_2$. All the above coordinate variables are implicitly assumed to be two-dimensional. Eq.~(\ref{quad}) \footnote{These equations have also been derived using the JIMWLK formalism by D. Triantafyllopoulos and by J. Jalilian-Marian (private communications).} suffers the same problem as Eq. (\ref{bk}) in the sense that it is not a closed equation because the right hand side includes higher-point correlations. The way to deal with this difficulty is the same as for the Balitsky-Kovchegov equation assuming that, for a large nucleus, these correlators can be factored as products of correlators involving only one trace at a time when the large-$N_c$ limit is taken. The resulting equation is equivalent to the quadrupole evolution equation found in Ref.~\cite{JalilianMarian:2004da} which was derived considering only the leading $N_c$ contributions from the beginning.  In other words, the full evolution equation has no terms which are explicitly suppressed by powers of $1/N_c$ and taking the large-$N_c$ limit only has the effect of allowing the aforementioned factorization of the correlators. This was also the case for the Balitsky-Kovchegov equation \cite{Kovchegov:1999yj} as can be seen from Eq. (\ref{bk}). 

Let us also comment on the physical interpretation of those seven terms on the right hand side of Eq.~(\ref{quad}). The first term stands for the virtual correction which subtracts probability from the original quadrupole. The second and third term represent the splitting of the quadrupole into two new dipoles while the last four terms indicate the splitting of the original quadrupole into a quadrupole and a new dipole. The kernels satisfy the relation Eq.~(\ref{sum}) as a result of conservation of probability.

Now, let us turn our attention to the unintegrated gluon distributions mentioned in the introduction. First consider the dipole gluon distribution whose operator definition is related to the dipole amplitude, namely, the two point functions of Wilson lines $\frac{1}{N_c}\left\langle\textrm{Tr}\left(U(x_{\perp})U^{\dagger}(y_{\perp})\right)\right\rangle$ as follows
\begin{equation}
xG^{(2)}(x,k_\perp)= \frac{q_{\perp }^{2}N_{c}}{2\pi^2 \alpha_s}%
\int d^2 x_{\perp}\int
\frac{d^2y_\perp}{(2\pi)^2}e^{-ik_\perp\cdot (x_\perp-y_\perp)}
\frac{1}{N_c}\left\langle\text{Tr}U(x_{\perp})U^\dagger(y_\perp)\right\rangle_{Y} .
\end{equation}
Its evolution is therefore provided by the Balitsky-Kovchegov equation.
 
The operator definition of the Weizs\"{a}cker-Williams gluon distribution is given by a slightly different operator which can be obtained from the quadrupole correlator. According to Ref.~\cite{Dominguez:2010xd, Dominguez:2011wm,Bomhof:2006dp}, the Weizs\"{a}cker-Williams gluon distribution can be written as 
\begin{equation}
xG^{(1)}(x,k_\perp)=-\frac{2}{\alpha_s}\int\frac{d^2v}{(2\pi)^2}\frac{d^2v'}{(2\pi)^2}\;e^{-ik_\perp\cdot(v-v')}\left\langle\text{Tr}\left[\partial_iU(v)\right]U^\dagger(v')\left[\partial_iU(v')\right]U^\dagger(v)\right\rangle_{Y}.
\end{equation}
The evolution equation\footnote{Inspired by Ref.~\cite{Metz:2011wb}, we find that the small-$x$ evolution equation of the Weizs\"{a}cker-Williams linearly polarized gluon distributions is given by a slightly different correlator $\left\langle\text{Tr}\left[\partial_iU(v)\right]U^\dagger(v')\left[\partial_j U(v')\right]U^\dagger(v)\right\rangle_{Y}$. We will leave the detailed derivation and phenomenological study of the linearly polarized gluon distributions to a future work \cite{lineargluon}.} for the correlator $\left\langle\text{Tr}\left[\partial_iU(v)\right]U^\dagger(v')\left[\partial_i U(v')\right]U^\dagger(v)\right\rangle_{Y}$ can be obtained from Eq.~(\ref{quad}) by differentiating with respect to $x_1$ and $x_2$, and then setting $x_1=x_2^\prime=v$ and $x_2=x_1^\prime=v^{\prime}$. Then the resulting evolution equation becomes
\begin{eqnarray}
&&\frac{\partial}{\partial Y}\left\langle\text{Tr}\left[\partial_iU(v)\right]U^\dagger(v')\left[\partial_iU(v')\right]U^\dagger(v)\right\rangle_{Y}\nonumber \\
&=&-\frac{\alpha_s N_c}{2\pi^2}\int d^2 z_{\perp} \frac{(v-v^\prime)^2}{(v-z)^2(z-v^\prime)^2}\left\langle\text{Tr}\left[\partial_iU(v)\right]U^\dagger(v')\left[\partial_iU(v')\right]U^\dagger(v)\right\rangle_{Y}\nonumber \\
&&-\frac{\alpha_s N_c}{2\pi^2}\int d^2 z_{\perp} \frac{1}{N_c}\frac{(v-v^\prime)^2}{(v-z)^2(z-v^\prime)^2}\left[\frac{(v-v^\prime)_i}{(v-v^\prime)^2}-\frac{(v-z)_i}{(v-z)^2}\right]\nonumber \\
&&\quad \times \left\{\left\langle\text{Tr}\left[U(v)U^\dagger(v')\left[\partial_iU(v')\right]U^\dagger(z)\right]\text{Tr}\left[U(z)U^\dagger(v)\right]\right\rangle_{Y}-\left\langle\text{Tr}\left[U(z)U^\dagger(v')\left[\partial_iU(v')\right]U^\dagger(v)\right]\text{Tr}\left[U(v)U^\dagger(z)\right]\right\rangle_{Y} \right\}\nonumber \\
&&-\frac{\alpha_s N_c}{2\pi^2}\int d^2 z_{\perp} \frac{1}{N_c}\frac{(v-v^\prime)^2}{(v-z)^2(z-v^\prime)^2}\left[\frac{(v^\prime-v)_i}{(v^\prime-v)^2}-\frac{(v^\prime-z)_i}{(v^\prime-z)^2}\right]\nonumber \\
&&\quad \times \left\{\left\langle\text{Tr}\left[\left[\partial_iU(v)\right]U^\dagger(z)U(v^\prime)U^\dagger(v)\right]\text{Tr}\left[U(z)U^\dagger(v^\prime)\right]\right\rangle_{Y}-\left\langle\text{Tr}\left[\left[\partial_iU(v)\right]U^\dagger(v^\prime)U(z)U^\dagger(v)\right]\text{Tr}\left[U(v^\prime)U^\dagger(z)\right]\right\rangle_{Y} \right\}\nonumber \\
&&-\frac{\alpha_s N_c}{\pi^2}\int d^2 z_{\perp} \frac{1}{N_c}\frac{1}{(v-z)^2(z-v^\prime)^2}\left[1-\frac{2\left((v-z)\cdot(z-v^\prime)\right)^2}{(v-z)^2(z-v^\prime)^2}\right]\nonumber \\
&&\quad \times \left\{\left\langle\textrm{Tr}\left[U(v^\prime)U^{\dagger}(z)\right]\textrm{Tr}\left[U(z)U^{\dagger}(v^\prime)\right]\right\rangle_{Y}+\left\langle\textrm{Tr}\left[U(v)U^{\dagger}(z)\right]\textrm{Tr}\left[U(z)U^{\dagger}(v)\right]\right\rangle_{Y}\right . \nonumber\\
&&\quad \quad \left .-\left\langle\textrm{Tr}\left[U(v^\prime)U^{\dagger}(v)\right]\textrm{Tr}\left[U(v)U^{\dagger}(v^\prime)\right]\right\rangle_{Y}-N_c^2\right\}.
\label{wwe}
\end{eqnarray}
Among these four terms in Eq.~(\ref{wwe}), the second and third terms are quite troublesome since they introduce new correlators involving three coordinates. The first term can be understood as the virtual correction as analogous to the first term in the Balitsky-Kovchegov equation shown in Eq.~(\ref{bk}). The last term is in agreement with the results obtained from the one-loop calculation in Ref.~\cite{Mueller:1999wm}. The calculation in Ref.~\cite{Mueller:1999wm} employs an effective theory which allows one to probe the Weizs\"{a}cker-Williams gluon distribution by using the current $j=-\frac{1}{4} F_{\mu\nu}^aF_{\mu\nu}^a$. One can also show that this calculation is equivalent to our above derivation for Eq.~(\ref{wwe}), and one can obtain the first three terms as well if all the graphs are included without angular average. 

\begin{figure}[tbp]
\begin{center}
\includegraphics[width=12cm]{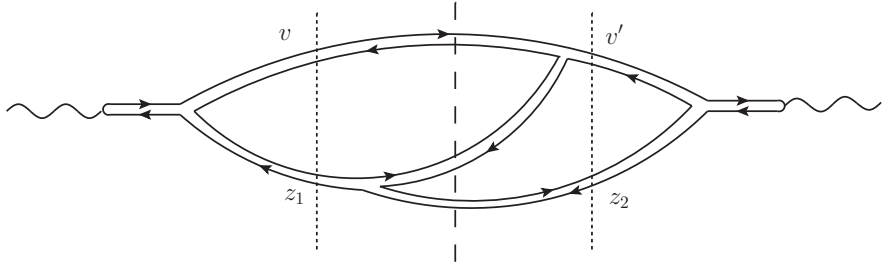}
\end{center}
\caption{\label{step2} Illustration of two-step evolution which generates the quadruple amplitude. The dotted lines indicate the moments of the interaction with the target nucleus and the dashed line represents the cut. The two dipoles correspond to the two internal color lines, and
are characterized by the coordinates $(z_1,v)$ and $(v',z_2)$ respectively
at the time of the interaction. The single external color line
interacts as a quadrupole defined by the coordinates
$(v,z_1,z_2,v')$.}
\end{figure}

In addition, we find that one will inevitably run into the evolution of quadrupoles irrespective of the initial conditions when the Weizs\"{a}cker-Williams gluon distribution appears in the process before small-$x$ evolution is included. As far as the Weizs\"{a}cker-Williams gluon distribution is concerned, one finds that after two steps of evolution, the contribution of quadrupoles appears as the following $S$-matrix amplitude
\begin{equation}
\frac{1}{N_c}\textrm{Tr}\left[U(z_1)U^{\dagger}(z_2)U(v^{\prime}) U^{\dagger}(v)\right]\frac{1}{N_c}\textrm{Tr}\left[U(v) U^{\dagger}(z_1)\right]\frac{1}{N_c}\textrm{Tr}\left[U(z_2)U^{\dagger}(v^{\prime})\right]\label{qplusd}
\end{equation}
Starting from the correlator $\left\langle\textrm{Tr}\left[\partial_iU(v)\right]U^\dagger(v')\left[\partial_i U(v')\right]U^\dagger(v)\right\rangle_{Y}$, the first step of evolution is given by Eq.~(\ref{wwe}) which generates terms like $\left\langle\text{Tr}\left[U(z_1)U^\dagger(v')\left[\partial_iU(v')\right]U^\dagger(v)\right]\text{Tr}\left[U(v)U^\dagger(z_1)\right]\right\rangle_{Y} $. One can further evolve such object and find that the second step of evolution yields a combination of a quadrupole plus two dipoles as in Eq.~(\ref{qplusd}). In terms of the dipole picture and 't Hooft's double line notation, we illustrate the above terms in Fig.~\ref{step2} where the double lines at $v$ and $v^{\prime}$ have the highest longitudinal momentum, the double line at $z_1$ has the next highest longitudinal momentum and the double line at $z_2$ has the smallest longitudinal momentum. Fig.~\ref{step2} and its radiative corrections are characterized by the feature that the double line in the middle does not directly connect to the virtual photons. By squaring the production amplitude, one gets the basic color and spatial structure of a quadrupole $\frac{1}{N_c}\textrm{Tr}\left[U(z_1)U^{\dagger}(z_2)U(v^{\prime}) U^{\dagger}(v)\right]$ together with two dipoles $\frac{1}{N_c}\textrm{Tr}\left[U(v) U^{\dagger}(z_1)\right]$ and $\frac{1}{N_c}\textrm{Tr}\left[U(z_2)U^{\dagger}(v^{\prime})\right]$ before further evolution. This shows that the Weizs\"acker-Williams distribution does not have a closed evolution equation on its own and, despite its apparently simpler structure in terms of only two coordinates, the full quadrupole evolution is needed to include small-$x$ effects. In terms of the DIS dijet process considered in \cite{Dominguez:2010xd, Dominguez:2011wm}, the correlation limit taken to avoid the quadrupole and be able to express the cross section in terms of the Weizs\"acker-Williams distribution does not help when small-$x$ evolution is considered. The same argument and conclusion also holds for the single gluon production in the gedanken DIS process using the current $j=-\frac{1}{4} F_{\mu\nu}^aF_{\mu\nu}^a$. However, for inclusive and semi-inclusive DIS, one can show that the evolution always involves only dipole amplitudes and this allows one to write a closed equation, namely the Balitsky-Kovchegov equation to describe these processes. 

\section{The weak interaction limit}

In order to study the evolution of color quadrupoles in the dilute regime, where the evolution will be described entirely by the BFKL evolution equation, let us consider the leading twist approximation to the dipole and quadrupole amplitudes and the corresponding evolution equations. In this leading twist approximation, the dipole amplitude $S(x_1,x_2)_Y=\frac{1}{N_c}\langle\text{Tr}U(x_1)U^\dagger(x_2)\rangle_Y$ can be written as $S(x_1,x_2)_Y=1-\frac{C_F}{2}\Gamma(x_1,x_2)_Y$ with $\Gamma$ satisfying the dipole form of the BFKL equation
\begin{equation}
\frac{\partial}{\partial Y}\Gamma(x_1,x_2)_Y=\frac{N_c\alpha_s}{2\pi^2}\int d^2z\,\frac{(x_1-x_2)^2}{(x_1-z)^2(x_2-z)^2}\left[\Gamma(x_1,z)_Y+\Gamma(z,x_2)_Y-\Gamma(x_1,x_2)_Y\right].\label{bfkl}
\end{equation}

For the quadrupole amplitude\footnote{Note that we have interchanged $x_2$ and $x_2^{\prime}$ in the definition of the quadrupole in order to match the physical observable calculated in the DIS dijet process.} $Q(x_1,x'_1,x'_2,x_2)_Y=\frac{1}{N_c}\langle\text{Tr}U(x_1)U^\dagger(x'_1)U(x'_2)U^\dagger(x_2)\rangle_Y$, we expand to second order in $gA$ with $A$ being the background gauge field. A simple calculation shows that this expansion can be written in terms of the linear dipole interaction $\Gamma$ (see Eq.~(B21) in Ref.~\cite{Dominguez:2011wm})\footnote{The BFKL limit has also been found by D. Triantafyllopoulos (private communication).}
\begin{equation}
Q(x_1,x'_1,x'_2,x_2)_Y=1-\frac{C_F}{2}\left[\Gamma(x_1,x_2)_Y+\Gamma(x'_1,x'_2)_Y+\Gamma(x_1,x'_1)_Y+\Gamma(x_2,x'_2)_Y-\Gamma(x_1,x'_2)_Y-\Gamma(x'_1,x_2)_Y\right].\label{qlinear}
\end{equation}
For the full quadrupole amplitude, all possible coordinate pairings have to be considered but it is important to remember that for the expression of the cross section for the dijet production in DIS, and therefore also for the derivation of the Weizs\"{a}cker-Williams gluon distribution, the quadrupole appears through the combination $Q(x_1,x'_1,x'_2,x_2)-S(x_1,x_2)-S(x'_1,x'_2)+1$. In the linear regime, the contributions from interactions hooking both gluons to the $(x_1,x_2)$ dipole or both gluons to the $(x'_1,x'_2)$ dipole cancel out.

Eq. (\ref{bfkl}) above can be obtained directly from the BK equation by expressing the dipole amplitude in terms of $\Gamma$ and neglecting the non-linear terms in $\Gamma$'s. The same approach can be easily followed for the quadrupole evolution equation. First, let us rewrite the quadrupole evolution equation in Eq.~(\ref{quad}) by grouping the different terms with the same dipole kernel
\begin{align}
&\frac{\partial}{\partial Y}Q(x_1,x'_1,x'_2,x_2)=-\frac{N_c\alpha_s}{(2\pi)^2}\int d^2z_\perp \notag\\
&\times\left\{\frac{(x_1-x_2)^2}{(x_1-z)^2(x_2-z)^2}\left[Q(x_1,x'_1,x'_2,x_2)-Q(z,x'_1,x'_2,x_2)S(x_1,z)-Q(x_1,x'_1,x'_2,z)S(x_2,z)+S(x_1,x_2)S(x'_1,x'_2)\right]\right.\notag\\
&+\frac{(x'_1-x'_2)^2}{(x'_1-z)^2(x'_2-z)^2}\left[Q(x_1,x'_1,x'_2,x_2)-Q(x_1,z,x'_2,x_2)S(x'_1,z)-Q(x_1,x'_1,z,x'_2)S(x'_2,z)+S(x_1,x_2)S(x'_1,x'_2)\right]\notag\\
&+\frac{(x_1-x'_1)^2}{(x_1-z)^2(x'_1-z)^2}\left[Q(x_1,x'_1,x'_2,x_2)-Q(z,x'_1,x'_2,x_2)S(x_1,z)-Q(x_1,z,x'_2,x_2)S(x'_1,z)+S(x_1,x'_1)S(x_2,x'_2)\right]\notag\\
&+\frac{(x_2-x'_2)^2}{(x_2-z)^2(x'_2-z)^2}\left[Q(x_1,x'_1,x'_2,x_2)-Q(x_1,x'_1,z,x_2)S(x'_2,z)-Q(x_1,x'_1,x'_2,z)S(x_2,z)+S(x_1,x'_1)S(x_2,x'_2)\right]\notag\\
&-\frac{(x_1-x'_2)^2}{(x_1-z)^2(x'_2-z)^2}\left[S(x_1,x_2)S(x'_1,x'_2)+S(x_1,x'_1)S(x_2,x'_2)-Q(x_1,z,x'_2,x_2)S(x'_1,z)-Q(x_1,x'_1,x'_2,z)S(x_2,z)\right]\notag\\
&\left.-\frac{(x'_1-x_2)^2}{(x'_1-z)^2(x_2-z)^2}\left[S(x_1,x_2)S(x'_1,x'_2)+S(x_1,x'_1)S(x_2,x'_2)-Q(z,x'_1,x'_2,x_2)S(x_1,z)-Q(x_1,x'_1,z,x_2)S(x'_2,z)\right]\right\},\label{fullquad}
\end{align}
where we have assumed that, in a large nucleus, one can factorize $\left\langle\mathcal{O}_1\mathcal{O}_2\right\rangle_Y$ into $\left\langle\mathcal{O}_1\right\rangle_Y\left\langle\mathcal{O}_2\right\rangle_Y$. Then we use the expressions quoted above for $Q$ and $S$ in terms of $\Gamma$ and keep only the linear terms. It is easy to see that each of the lines of Eq. (\ref{fullquad}) becomes a BFKL equation of the form given in Eq. (\ref{bfkl}) for each pair of coordinates. This is in complete agreement with what one would expect from just evolving Eq. (\ref{qlinear}) with BFKL in the linear regime. BFKL evolution generates the exponential growth of the quadrupole $T$-matrix ($T_Q=1-Q$) in terms of the rapidity $Y$. The saturation scale $Q_s(Y)$ is usually used to characterize the scale at which $T$ approaches $1$. If one fixes the relative size of all the spatial coordinates of the quadrupole, one expects that quadrupole evolution in the dilute regime should obey one single BFKL equation and thus exhibit the same geometrical scaling behavior as for dipoles \cite{Stasto:2000er,GolecBiernat:2001if,Iancu:2002tr} in terms of the variable $r^2Q_s^2(Y)$ with $r$ being the typical scale of the quadrupole. 

Similarly, in the dilute regime, the correlator involved in the calculation of the Weizs\"acker-Williams distribution can be written in terms of $\Gamma$'s as:
\begin{equation}
\left\langle\text{Tr}\left[\partial_iU(v)\right]U^\dagger(v')\left[\partial_iU(v')\right]U^\dagger(v)\right\rangle_{Y}=\frac{C_F}{2}\partial_{v_i}\partial_{v'_i}\Gamma(v,v').\label{linearcorrww}
\end{equation}
Naturally, expanding the correlators in Eq. (\ref{wwe}) and keeping only terms linear in $\Gamma$ or its derivatives leads to the same result as differentiating twice Eq. (\ref{bfkl}).

In the dilute regime where the gluon density is low, we know that the Weizs\"{a}cker-Williams gluon distribution $xG^{(1)}$ and the dipole gluon distribution $xG^{(2)}$ both reduce to the same leading twist result. To see this connection explicitly recall that $xG^{(1)}$ is given by the Fourier transform of the correlator in (\ref{linearcorrww}) and therefore the derivatives can be replaced by $k_\perp^2$ after integrating by parts.

\section{The Unitarity limit and fixed points}
Let us begin with the discussion on the dipole amplitude. Schematically, the Balitsky-Kovchegov equation can be cast into
\begin{equation}
\frac{\partial}{\partial Y} S=\int P_{d\to dd} \left(S S-S\right),\label{bk2}
\end{equation}
with $S$ being the dipole amplitude and $P_{d\to dd} >0 $ being the probability of a dipole splitting into two new dipoles per unit rapidity.

With Eq.~(\ref{sum}) and the assumption of factorization of operators, the evolution equation of quadrupoles can be schematically written as 
\begin{equation}
\frac{\partial}{\partial Y} Q=\int P_{q\to qd} \left(Q S-Q\right)+\int  P_{q\to dd} \left(S S-Q\right),\label{quads}
\end{equation}
where $Q$ represents the generic quadrupole amplitude, $P_{q\to qd}$ is the probability that the original quadrupole splits into a quadrupole and a dipole due to gluon emissions and $P_{q\to dd}$ stands for the probability of the original quadrupole splitting into two dipoles. The probabilistic interpretation of the quadrupole evolution equation is of course only heuristic since $P_{q\to qd}$ and $P_{q\to dd}$ are not positive-definite. However, we find that $P_{q\to qd}+P_{q\to dd}= \frac{\alpha_s N_c d^2 z}{(2\pi)^2}  \mathcal{K}_1$ is always positive (this equality is the identity in Eq.~(\ref{sum})). It seems to make sense to speak of the probability that the original quadrupole decays to $qd$ and $dd$ inclusively, but not that it decays to $qd$ or $dd$ exclusively. 

$S=0$ and $S=1$ are two fixed points of Eq.~(\ref{bk2}), and $Q=0$ and $Q=1$ are fixed points of Eq.~(\ref{quads}) with $S=Q=0$ a stable fixed point and $S=Q=1$ an unstable fixed point. Let us first write $S=0+\delta S$, $Q=0+\delta Q$ where $\delta S$ and $\delta Q$ are
small perturbations, then Eqs.~(\ref{bk2}) and (\ref{quads}) become $\frac{\partial}{\partial Y} \delta S=-\int P_{d\to dd} \delta S$ and $\frac{\partial}{\partial Y} \delta Q=-\int  \left(P_{q\to qd}+ P_{q\to dd}\right) \delta Q$. Since $P_{d\to dd} >0 $ and $P_{q\to qd}+P_{q\to dd}\propto \mathcal{K}_1 >0$, the evolution equations will drive $\delta S$ and $\delta Q$ to zero, thus $S=Q=0$ is a stable fixed point. Now let us take $S=1-\delta S$ and $Q=1-\delta Q$, then Eq.~(\ref{bk2}) becomes $\frac{\partial}{\partial Y} \delta S=\int P_{d\to dd}(x,y,z) \left(\delta S(x,z)+\delta S(y,z)-\delta S(x,y)\right)$ which is exactly the BFKL equation. Similarly, as shown in Sec. III, $\delta Q$ obeys the BFKL equation in the weak interaction limit as well. It is well-known that BFKL evolution will drive $\delta S$ and $\delta Q$ away from zero in terms of exponential growth. Therefore, the fixed point $S=Q=1$ is unstable. We do not know if these fixed points are the only fixed points for the quadrupole evolution, but it is certainly true for the Balitsky-Kovchegov equation. As a result, we expect that the evolution will always drive the dipole and quadrupole amplitudes from the dilute limit towards the stable fixed point $S=Q=0$ which leads to unitarity and saturation. 
 
Finally, we note that there is a traveling wave picture \cite{Braun:2000wr,Munier:2003vc} for the evolution of $Q$, or of $T_Q =1-Q$, exactly analogous to that for $S$, or $T=1-S$. The velocities of the traveling waves for $T_Q$ and $T$ are identical, since the velocity is determined by BFKL evolution. If one scales all coordinates in $T_Q$ uniformly then the shapes of the traveling wave front of $T_Q$ and $T$ are identical except near the top of the fronts, that is where nonlinear terms in the evolution become important. 

\section{Conclusion}
In summary, using the JIMWLK Hamiltonian, we have derived the evolution equation for the quadrupole and the Weizs\"{a}cker-Williams gluon distribution at finite $N_c$. We find that they follow BFKL evolution in the dilute regime and reach the saturation regime as a stable fixed point. Following the discussion in Ref.~\cite{Munier:2003vc,Mueller:2002zm}, we know that BFKL evolution together with a saturation boundary are responsible for the geometrical scaling behavior \cite{Stasto:2000er,GolecBiernat:2001if,Iancu:2002tr} of the dipole gluon distribution. Since we also observe the same properties for the quadrupole evolution equation, we believe that the Weizs\"{a}cker-Williams gluon distribution should exhibit geometrical scaling behavior as well, although its evolution equation is much more complicated. It seems that quadrupoles evolve essentially the way dipoles do, but their evolution is more difficult to evaluate due to the complicated structure of the evolution equation. The difference of their small-$x$ evolution behavior lies in the transition region between the scaling regime and the saturation regime. Likely, numerical studies of quadrupole evolution will be necessary to understand the details of their different behavior. 

\begin{acknowledgements}
We thank Giovanni Chirilli, Jamal Jalilian-Marian, Jianwei Qiu, Dionysis Triantafyllopoulos and Feng Yuan for helpful discussions. This work was supported in part by the U.S.
Department of Energy and by the DOE OJI grant No. DE - SC0002145. The work of SM was partly supported by the Agence Nationale de la Recherche 
(France), contract ANR-06-JCJC-0084-02.
\end{acknowledgements}

\appendix
\section{Derivation of the evolution equation for the quadrupole amplitude}
\label{derivation}
Here we present some essential details of the derivation leading to Eq.~(\ref{quad}). We begin with Eq.~(\ref{JH}) with the operator $\mathcal{O}$ set to be $\frac{1}{N_c}\left\langle\textrm{Tr}\left(U(x_{1})U^{\dagger}(x^{\prime}_{1})U(x_{2})U^{\dagger}(x^{\prime}_{2})\right)\right\rangle_{Y}$. Since there are four different terms inside the square bracket in the definition of $\eta_{uv}^{ab}$ as shown in Eq.~(\ref{JK}), we compute the contributions of these four terms separately and provide the results as follows.

The first term which involves the identity matrix in color space $\delta^{ab}$ yields the contribution $I_1$ 
\begin{eqnarray}
I_1 &=&-\frac{\alpha_s C_F}{2\pi^2}\int d^2 z_{\perp} \left[\frac{(x^{\prime}_1-x_1)^2}{(x^{\prime}_1-z)^2(z-x_1)^2}+\frac{(x^{\prime}_2-x_2)^2}{(x^{\prime}_2-z)^2(z-x_2)^2}\right]\left\langle\textrm{Tr}\left[U(x_{1})U^{\dagger}(x^{\prime}_{1})U(x_{2})U^{\dagger}(x^{\prime}_{2})\right]\right\rangle_{Y}\nonumber \\
&&+\frac{\alpha_s N_c }{(2\pi)^2}\int d^2 z_{\perp} \mathcal{A}(x_1, x_1^{\prime},x_2, x_2^{\prime};z)
\left\{\frac{1}{N_c}\left\langle\textrm{Tr}\left[U^{\dagger}(x^{\prime}_{1})U(x_{2})\right]\textrm{Tr}\left[U^{\dagger}(x^{\prime}_{2})U(x_{1})\right]\right\rangle_{Y}\right. \nonumber \\
&& \quad \quad\left . -\frac{1}{N_c^2}\left\langle\textrm{Tr}\left[U(x_{1})U^{\dagger}(x^{\prime}_{1})U(x_{2})U^{\dagger}(x^{\prime}_{2})\right]\right\rangle_{Y}\right\} .\label{q1}
\end{eqnarray}
With the help of the identity $T^b\tilde{V}_{v}^{ba}=U_vT^a U^{\dagger}_v$, the second term which involves the term $\left(\tilde{V}^{\dagger}_{u}\tilde{V}_{v}\right)^{ab}$ gives the contribution $I_2$
\begin{eqnarray}
I_2 &=&-\frac{\alpha_s C_F}{2\pi^2}\int d^2 z_{\perp} \left[\frac{(x^{\prime}_1-x_2)^2}{(x^{\prime}_1-z)^2(z-x_2)^2}+\frac{(x^{\prime}_2-x_1)^2}{(x^{\prime}_2-z)^2(z-x_1)^2}\right]\left\langle\textrm{Tr}\left[U(x_{1})U^{\dagger}(x^{\prime}_{1})U(x_{2})U^{\dagger}(x^{\prime}_{2})\right]\right\rangle_{Y}\nonumber \\
&&+\frac{\alpha_s N_c }{(2\pi)^2}\int d^2 z_{\perp} \mathcal{B}(x_1, x_1^{\prime},x_2, x_2^{\prime};z)
\left\{\frac{1}{N_c}\left\langle\textrm{Tr}\left[U(x_{1})U^{\dagger}(x^{\prime}_{1})\right]\textrm{Tr}\left[U(x_{2})U^{\dagger}(x^{\prime}_{2})\right]\right\rangle_{Y}\right. \nonumber \\
&& \quad \quad\left . -\frac{1}{N_c^2}\left\langle\textrm{Tr}\left[U(x_{1})U^{\dagger}(x^{\prime}_{1})U(x_{2})U^{\dagger}(x^{\prime}_{2})\right]\right\rangle_{Y}\right\} .\label{q2}
\end{eqnarray}
The third term $-\left(\tilde{V}^{\dagger}_{u}\tilde{V}_{z}\right)^{ab}$ and the fourth term $-\left(\tilde{V}^{\dagger}_{z}\tilde{V}_{v}\right)^{ab}$ give the contributions $I_3$ and $I_4$ respectively. These have identical forms but different variables, their sum reads
\begin{eqnarray}
&&I_3+I_4\nonumber \\
&=&\frac{\alpha_s N_c}{(2\pi)^2}\int d^2 z_{\perp} \mathcal{K}_2(x_1; x_1^{\prime}, x_2^{\prime};z)\frac{1}{N_c}\left\langle\textrm{Tr}\left[U(x_{1})U^{\dagger}(z)\right]\textrm{Tr}\left[U(z)U^{\dagger}(x^{\prime}_{1})U(x_{2})U^{\dagger}(x^{\prime}_{2})\right]\right\rangle_{Y}\nonumber \\
&&+\frac{\alpha_s N_c}{(2\pi)^2}\int d^2 z_{\perp} \mathcal{K}_2(x_1^{\prime}; x_1, x_2;z)\frac{1}{N_c}\left\langle\textrm{Tr}\left[U(z)U^{\dagger}(x^{\prime}_1)\right]\textrm{Tr}\left[U(x_1)U^{\dagger}(z)U(x_{2})U^{\dagger}(x^{\prime}_{2})\right]\right\rangle_{Y}\nonumber \\
&&+\frac{\alpha_s N_c}{(2\pi)^2}\int d^2 z_{\perp} \mathcal{K}_2(x_2; x_1^{\prime}, x_2^{\prime};z)\frac{1}{N_c}\left\langle\textrm{Tr}\left[U(x_2)U^{\dagger}(z)\right]\textrm{Tr}\left[U(x_1)U^{\dagger}(x_1^{\prime})U(z)U^{\dagger}(x^{\prime}_{2})\right]\right\rangle_{Y}\nonumber \\
&&+\frac{\alpha_s N_c}{(2\pi)^2}\int d^2 z_{\perp} \mathcal{K}_2(x_2^{\prime}; x_1, x_2;z)\frac{1}{N_c}\left\langle\textrm{Tr}\left[U(z)U^{\dagger}(x^{\prime}_2)\right]\textrm{Tr}\left[U(x_1)U^{\dagger}(x_1^{\prime})U(x_{2})U^{\dagger}(z)\right]\right\rangle_{Y}\nonumber \\
&&-\frac{\alpha_s N_c}{(2\pi)^2}\int d^2 z_{\perp} \frac{1}{N_c^2}\left\langle\textrm{Tr}\left[U(x_{1})U^{\dagger}(x^{\prime}_{1})U(x_{2})U^{\dagger}(x^{\prime}_{2})\right]\right\rangle_{Y}\nonumber \\
&&\quad \times\left[\mathcal{K}_2(x_1; x_1^{\prime}, x_2^{\prime};z)+\mathcal{K}_2(x_1^{\prime}; x_1, x_2;z)+\mathcal{K}_2(x_2; x_1^{\prime}, x_2^{\prime};z)+\mathcal{K}_2(x_2^{\prime}; x_1, x_2;z)\right] .\label{q34}
\end{eqnarray}
Noticing that $C_F=\frac{N_c^2-1}{2N_c}$ and using Eq.~(\ref{sum}), it is straightforward to check that the coefficient of the non-leading $N_c$ terms $\frac{1}{N_c^2}\left\langle\textrm{Tr}\left[U(x_{1})U^{\dagger}(x^{\prime}_{1})U(x_{2})U^{\dagger}(x^{\prime}_{2})\right]\right\rangle_{Y}$ vanishes. 
In the end, we find the sum of all these four contributions leads to the right hand side of Eq.~(\ref{quad}) without finite $N_c$ corrections.

\end{document}